\documentclass[12pt]{article}
\usepackage{epsfig,feynmf,amssymb,slashed}

\def\beq{\begin{equation}}
\def\eeq#1{\label{#1}\end{equation}}
\def\eeqn{\end{equation}}

\def\beqa{\begin{eqnarray}}
\def\eeqa#1{\label{#1}\end{eqnarray}}
\def\eeqan{\end{eqnarray}}

\let\bar=\overbar

\def\Dslash{\not{\hbox{\kern-4pt $D$}}}
\def\dslash{\not{\hbox{\kern-2pt $\del$}}}

\def\msb{{\bar{\ssstyle M \kern -1pt S}}}

\def\Title#1{\begin{center} {\Large {\bf #1} } \end{center}}

\begin{document}

\Title{Redesigning Electroweak Theory: Does the Higgs Particle Exist?\footnote{Talk given by JWM at the Tenth Workshop on Non-Perturbative QCD at l'Institut d'Astrophysique de Paris, France, 8--12 June 2009}}

\bigskip\bigskip


\begin{raggedright}

{\it J.~W. Moffat$^{1,2}$\index{Moffat, J.~W.} and V.~T. Toth$^1$\index{Toth, V.~T.}\\
$^1$Perimeter Institute for Theoretical Physics, Waterloo, Ontario N2L 2Y5, Canada\\
$^2$Department of Physics, University of Waterloo, Waterloo, Ontario N2L 3G1, Canada}
\bigskip\bigskip
\end{raggedright}

\begin{abstract}
An electroweak model in which the masses of the W and Z bosons and the fermions are generated by quantum loop graphs through a symmetry breaking is investigated. The model is based on a regularized quantum field theory in which the quantum loop graphs are finite to all orders of perturbation theory and the massless theory is gauge invariant, Poincar\'e invariant, and unitary. The breaking of the electroweak symmetry SU(2)$_L\times$U(1)$_Y$ is achieved without a Higgs particle. A fundamental energy scale $\Lambda_W\simeq 542$~GeV (not to be confused with a naive cutoff) enters the theory through the regularization of the Feynman loop diagrams. The theory yields a prediction for the W mass $m_W$ that is accurate to about 0.5\% without radiative corrections. The scattering amplitudes for W$^+_L$W$^-_L\rightarrow$W$^+_L$W$^-_L$ and e$^+$e$^-\rightarrow$W$^+_L$W$^-_L$ processes do not violate unitarity at high energies due to the suppression of the amplitudes by the running of the coupling constants at vertices.
\end{abstract}

\begin{fmffile}{fewfigs}

\section{Introduction}

Despite its phenomenological success, theoretical problems prompt searching for an alternative to the standard electroweak (EW) theory. There is the serious Higgs hierarchy problem (unstable Higgs mass) and the cosmological constant problem. {\bf The origin of the symmetry breaking mechanism remains elusive after almost 50 years.}

In this paper, we review an alternative to the standard model \cite{Moffat1990,Evens1991,Moffat1991,Clayton1991b}, in the form of a theory that employs a nonlocal regularization scheme and symmetry breaking through the path integral measure. The foundations of this theory are described briefly in Section~\ref{sec:found}. In Section~\ref{sec:appl} we apply the theory and calculate vector boson masses and the $\rho$ parameter, demonstrate how the theory can be used to obtain fermion masses, and also show that, through the running of the coupling constants $g$ and $g'$, the theory achieves unitarity.

\section{Foundations}
\label{sec:found}

The standard EW model gains mass for the W and Z bosons, while keeping the photon massless by introducing a {\bf classical} scalar field into the action. This scalar degree of freedom is assumed to transform as an isospin doublet, spontaneously breaking the SU(2)$_L\times$U(1)$_Y$ by a Higgs mechanism at the purely classical tree graph level.

The standard and commonly accepted explanation for the origin of this mechanism is a spontaneous symmetry breaking framework in which the symmetry SU(2)$_L\times$U(1)$_Y$ is not broken by the interactions but is ``softly'' broken by the asymmetry of the ground state (vacuum state).

\subsection{The standard electroweak model with a Higgs particle}

The spontaneous breaking of $\mathrm{SU}(2)_L\times\mathrm{U}(1)_Y\rightarrow\mathrm{U}(1)_\mathrm{em}$ generates a value for the vacuum density, which is some $10^{56}$ times larger in magnitude than the observed value $\rho^\mathrm{vac}_\mathrm{obs}\sim(0.0024~\mathrm{eV})^4$ and has the wrong sign. This is known as the cosmological constant problem.

The tree-level (bare) Higgs mass receives large quadratically-divergent corrections from the Higgs loop diagrams. This is the Higgs mass hierarchy problem.

Discovering a satisfactory alternative has proved to be {\bf highly nontrivial}. Proposed alternatives face severe problems. New particle contributions at less than 1 or 2 TeV level can affect precision EW data that can generate unacceptably large effects; significant fine tuning may be required at least at the 1-percent level. These models include MSSM, Little Higgs, pseudo-Goldstone bosons. Extensions of the standard EW model such as technicolor, and other composite models can face unacceptably large flavor changing contributions and CP violation.

For a classical potential $V(\phi)$ for a scalar field $\phi$ we can identify: $\left<T_{\mu\nu}\right>_0=V(\phi)=\rho_\mathrm{vac}$, where $T_{\mu\nu}$ is the energy-momentum tensor and $\rho_\mathrm{vac}$ is the vacuum energy density. The Higgs field vacuum energy is calculated from the classical Higgs potential:
\begin{equation}
V(\phi)=V_0-\mu^2\phi^2+\lambda\phi^4,~~~~~\left<0|\phi|0\right>=v,
\end{equation}
where $v\sim 250$~GeV is the EW energy scale. We have $\mu^4=\lambda^2v^4$. From the minimization of the Higgs potential we obtain $\phi_\mathrm{min}=\mu^2/2\lambda$ and $V_\mathrm{min}=V_0-\mu^4/4\lambda=\rho_\mathrm{vac}^\mathrm{ssb}$ for the spontaneous symmetry breaking vacuum energy density. Choosing $V(0)=0$ we obtain
\begin{equation}
\rho_\mathrm{vac}^\mathrm{ssb}=-\frac{\mu^4}{4\lambda}\sim-4\lambda v^4\sim-10^{5}~\mathrm{GeV}^4,~~~~~\rho_\mathrm{vac}^\mathrm{obs}\sim 10^{-47}~\mathrm{GeV}^4,~~~~~|\rho_\mathrm{vac}^\mathrm{ssb}|\sim 10^{56}\rho_\mathrm{vac}^\mathrm{obs}.
\end{equation}

The minimal EW model with a Higgs doublet is consistent with the experimental bounds on flavor changing neutral currents and CP violation.

The primary target of an EW global fit is the prediction of the Higgs mass $M_\mathrm{H}$. The complete fit represents the most accurate estimation of $M_\mathrm{H}$ considering all available data \cite{Flacher:2008zq}. The result is $M_\mathrm{H}=116.4^{+18.3}_{-1.3}$~GeV where the error accounts for both experimental and theoretical uncertainties. The result for the standard fit without the direct Higgs searches is $M_\mathrm{H}=80^{+30}_{-23}$~GeV and the 2$\sigma$ and 3$\sigma$ intervals are, respectively, $[39,155]$~GeV and $[26,209]$~GeV.

We conclude from this that the minimal EW model requires a light Higgs.

\subsection{The gauge invariant local EW theory}

The theory is based on the local SU(2)$_L\times$U(1)$_Y$ invariant Lagrangian that includes leptons and quarks (with the color degree of freedom of the strong interaction group SU(3)$_c$) and the boson vector fields that arise from gauging the $\mathrm{SU}(2)_L\times\mathrm{U}(1)_Y$ symmetry:
\begin{equation}
L_\mathrm{local}=L_F+L_W+L_B+L_I.
\end{equation}
$L_F$ is the free fermion Lagrangian consisting of massless kinetic terms for each fermion:
\begin{eqnarray}
L_F=\sum_\psi\bar\psi i\slashed\partial\psi=\sum_{q^L}\bar q^Li\slashed\partial q^L+\sum_f\bar\psi^Ri\slashed\partial\psi^R,&~~&q^L\in\left[\left(\matrix{\nu^L\cr e^L}\right),\left(\matrix{u^L\cr d^L}\right)_{r,g,b}\right],
\end{eqnarray}
\vskip -24pt
\begin{eqnarray}
L_B=-\frac{1}{4}B^{\mu\nu}B_{\mu\nu},&~~~~~&B_{\mu\nu}=\partial_\mu B_\nu-\partial_\nu B_\mu,\\
L_W=-\frac{1}{4}W_{\mu\nu}^aW^{a\mu\nu},&~~~~~&W_{\mu\nu}^a=\partial_\mu W_\nu^a-\partial_\nu W_\mu^a-gf^{abc}W_\mu^bW_\nu^c,
\end{eqnarray}
where we used the metric convention $\eta_{\mu\nu}=\mathrm{diag}(+1,-1,-1,-1)$, and set $\hbar=c=1$.
{\bf Note that there is no classical scalar field contribution in the Lagrangian.}

The SU(2) generators satisfy
\begin{equation}
[T^a,T^b]=if^{abc}T^c,~~~~~\mathrm{with}~~~~~T^a=\frac{1}{2}\sigma^a.
\end{equation}

The fermion gauge-boson interaction terms are
\begin{equation}
L_I=-gJ^{a\mu}W_\mu^a-g'J_Y^\mu B_\mu,~~~~~J^{a\mu}=\sum_{q^L}\bar q^L\gamma^\mu T^aq^L,~~~~~\mathrm{and}~~~~~J_Y^\mu=\sum_\psi\frac{1}{2}Y_\psi\bar\psi\gamma^\mu\psi,
\end{equation}
\vskip -24pt
\begin{eqnarray}
Y(e^L)=Y(\nu^L)=-1,~~~~~Y(e^R)=-2,~~~~~Y(\nu^R)=0,\nonumber\\
Y(u^L)=Y(d^L)=\frac{1}{3},~~~~~Y(u^R)=\frac{4}{3},~~~~~Y(d^R)=\frac{2}{3}.
\end{eqnarray}

$L$ is invariant under the local gauge transformations (order $g,g'$):
\begin{eqnarray}
\delta\psi^L=-\left(igT^a\theta^a+ig'\frac{Y_\psi}{2}\beta\right)\psi^L,&~~~~~&\delta\psi^R=-ig'\frac{Y_\psi}{2}\beta\psi^R,\\
\delta W_\mu^a=\partial_\mu\theta^a+gf^{abc}\theta^bW_\mu^c,&~~~~~&\delta B_\mu=\partial_\mu B.
\end{eqnarray}
$L$ is an SU(2)$_L\times$U(1)$_Y$ invariant Lagrangian.

Quantization is accomplished via the path integral formalism:
\begin{equation}
\left<T({\cal O}[\phi])\right>\propto\int[D\bar\psi][D\psi][DW][DB]\mu_\mathrm{inv}[\bar\psi,\psi,B,W]{\cal O}[\phi]e^{i\int d^4xL_\mathrm{local}}.
\end{equation}
In the local case, the invariant integration measure $\mu_\mathrm{inv}$ is the trivial one.

We have to gauge fix the Lagrangian:
\begin{equation}
L_\mathrm{GF}=-\frac{1}{2\xi}(\partial_\mu B^\mu)^2-\frac{1}{2\xi}(\partial_\mu W^{a\mu})^2.
\end{equation}

We look at diagonalizing the charged sector and mixing in the neutral boson sector. If we write
\begin{equation}
W^\pm=\frac{1}{\sqrt{2}}(W^1\mp iW^2),
\end{equation}
then we get the fermion interaction terms
\begin{equation}
-\frac{g}{\sqrt{2}}(J_\mu^+W^{+\mu}+J_\mu^-W^{-\mu}),
\end{equation}
where the charged current is given by
\begin{equation}
J_\mu^\pm=J_{1\mu}^\pm\pm iJ_{2\mu}^\pm=\sum_{q_L}\bar q^L\gamma_\mu T^\pm q^L~~~~~{\rm implying}~~~~~J_\mu^+=\sum_{q_L}(\bar\nu^L\gamma_\mu e^L+\bar u^L\gamma_\mu d^L).
\end{equation}

In the neutral sector, we can mix the fields in the usual way:
\begin{equation}
Z_\mu=c_wW_\mu^3-s_wB_\mu~~~\mathrm{and}~~~A_\mu=c_wB_\mu+s_wW_\mu^3,
\label{eq:2.35}
\end{equation}
where $s_w=\sin\theta_w$ and $c_w=\cos\theta_w$ with $\theta_w$ denoting the weak mixing (Weinberg) angle. We define the usual trigonometric relations
\begin{equation}
s^2_w=\frac{g'^2}{g^2+g'^2}~~~\mathrm{and}~~~c^2_w=\frac{g^2}{g^2+g'^2}.
\end{equation}
The neutral current fermion interaction terms now look like:
\begin{equation}
-gJ^{3\mu}W_\mu^3-g'J_Y^\mu B_\mu=-(gs_wJ^{3\mu}+g'c_wJ_Y^\mu)A_\mu-(gc_wJ^{3\mu}-g's_wJ_Y^\mu)Z_\mu.
\end{equation}

If we identify the resulting $A_\mu$ field with the photon, then we have the unification condition:
\begin{equation}
e=gs_w=g'c_w
\end{equation}
and the electromagnetic current is
\begin{equation}
J_\mathrm{em}^\mu=J^{3\mu}+J_Y^\mu,
\end{equation}
where $e$ is the charge of the proton. Note that the coupling now looks like $(Q-T^3)+T^3=Q$ and we only get coupling of the photon to charged fermions at tree level. We can then identify the neutral current:
\begin{equation}
J_\mathrm{NC}^\mu=J^{3\mu}-s_wJ_\mathrm{em}^\mu,
\end{equation}
and write the fermion-boson interaction terms as
\begin{equation}
L_I=-\frac{g}{\sqrt{2}}(J_\mu^+W^{+\mu}+J_\mu^-W^{-\mu})-gs_wJ_\mathrm{em}^\mu A_\mu-\frac{g}{c_w}J_\mathrm{NC}^\mu Z_\mu.
\end{equation}
This, along with the suitably rewritten boson interaction terms, gives the usual vertices of the local point theory.

\subsection{The gauge invariant regularized theory}

To regularize the fields, we write the non-local (smeared) fields as a convolution of the local fields with a function whose momentum space Fourier transform is an {\bf entire} function. This function can be related to a Lorentz invariant operator distribution as~\cite{Moffat1990,Evens1991,Moffat2007f,Moffat2008b,Moffat2008c}:
\begin{equation}
\Phi(x)=\int d^4yG(x-y)\phi(y)=G\biggl(\frac{\Box}{\Lambda_W^2}\biggr)\phi(x),
\end{equation}
where $\phi(x)$ is a local field and $\Lambda_W$ denotes a non-local electroweak energy scale. We make a choice of a specific smearing operator:
\begin{equation}
G\left(\frac{\Box}{\Lambda_W^2}\right)\equiv{\cal E}_m=\exp\left(-\frac{\Box+m^2}{2\Lambda_W^2}\right).
\end{equation}
Since the theory is initially massless, all fields are smeared with ${\cal E}_0$. We now write the initial
Lagrangian in non-local form:
\begin{equation}
L_\mathrm{reg}=L[\phi]_F+{\cal L}[\Phi]_I,
\end{equation}
where ${\cal L}[\Phi]_I$ indicates smearing of the interacting fields.

An essential feature of the regularized, non-local field theory is the requirement that the classical tree graph theory remains local, giving us a well defined classical limit in the gauge invariant case \cite{Evens1991}.

We first note that we must alter the quantized form of the theory by generalizing the path integral~\cite{Evens1991}:
\begin{equation}
\left<T^*({\cal O}[\Phi])\right>\propto\int[D\bar\psi][d\psi][DW][DB][D\bar\eta][D\eta][D\bar c][Dc]{\cal O}
[\Phi]\exp(iS_0[\phi]+iS_I[\Phi]),
\end{equation}
where $\eta$ and $c$ are ghost fields, and we are now dealing with expectation values of operators that are functionals of the smeared fields $\Phi$.

To generate a perturbation scheme in the non-local operators, we write the generating functional as
\begin{equation}
W[{\cal J}]=\ln(Z[{\cal J}])=\ln\left(\int[D\phi]\exp\left(i\int dx\{L_F[\phi]+{\cal L}_I[\Phi]
+{\cal J}(x)\Phi(x)\}\right)\right),
\end{equation}
where the source term ${\cal J}$ is now non-local. We note that in momentum space, the smeared fields are related one-to-one to the local fields:
\begin{equation}
\Phi(p)=G(p^2)\phi(p)=\exp\left(\frac{p^2-m^2}{2\Lambda_W^2}\right)\phi(p).
\end{equation}

\subsection{Breaking the symmetry with a path integral measure}

We break SU(2)$_L\times$U(1)$_Y$ down to U(1)$_\mathrm{em}$ {\bf not at the classical level} as is done in the standard model, which generates boson masses at tree level, but in the quantum regime, so that all the effects show up at loop order (which is where the non-locality shows up as well, as both are quantum effects). This means {\bf leaving the action gauge invariant and modifying the measure}, which alters the quantization of the theory, in order to produce the desired results.

The symmetry breaking measure in our path integral generates three new degrees of freedom as scalar Nambu-Goldstone bosons that give the W$^\pm$ and Z$^0$ bosons longitudinal modes, which makes them massive while retaining a massless photon.

Since we want to mix the W$_3$ and B to get massive W$^\pm$ and Z$^0$ bosons and a massless photon, we need to work with the measure in a sector which is common to all gauge bosons. This implies working with the fermion contributions and leaving the bosonic and ghost contributions invariant.

The self-energy contribution coming from

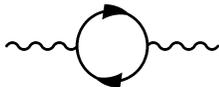
\begin{center}
\begin{fmfgraph*}(80,40)
\fmfleftn{i}{1}\fmfrightn{o}{1}
\fmf{boson}{i1,v1}
\fmf{boson}{v2,o1}
\fmf{fermion,left,tension=0.5}{v1,v2,v1}
\end{fmfgraph*}
\end{center}
\vskip -12pt
\noindent is given by
\begin{eqnarray}
-i\Pi_f^L&=&-\frac{4iee'\Lambda_W^2}{(4\pi)^2}[g_+(K_{m_1m_2}-L_{m_1m_2})+g_-M_{m_1m_2}],\\
-\Pi_f^T&=&-\frac{4iee'\Lambda_W^2}{(4\pi)^2}[g_+(K_{m_1m_2}-L_{m_1m_2}+2P_{m_1m_2})+g_-M_{m_1m_2}],
\end{eqnarray}
where $K_{m_1m_2}$, $L_{m_1m_2}$, $M_{m_1m_2}$ and $P_{m_1m_2}$ are all functions of the masses $m_1,m_2$ and the Euclidean momentum $p_E$ \cite{Moffat2008b}, and where
\begin{equation}
f_{m_1m_2}=\frac{m_1^2}{\Lambda_W^2}+\frac{\tau}{1-\tau}\frac{m_2^2}{\Lambda_W^2}.
\end{equation}
If we insert this into the quadratic terms in the action and invert, we get the corrected propagator (in a general gauge):
\begin{equation}
iD^{\mu\nu}=-i\left(\frac{\eta^{\mu\nu}-\frac{p^\mu p^\nu}{p^2}}{p^2-\Pi_f^T}+\frac{\frac{\xi p^\mu p^\nu}{p^2}}{p^2-\xi\Pi_f^L}\right).
\label{eq:vecprop}
\end{equation}

When the longitudinal piece $\Pi_L$ is nonzero in the unitary gauge (where only the physical particle spectrum remains), we have no unphysical poles in the longitudinal sector. In this way, we can assure ourselves that we are not introducing spurious degrees of freedom into the theory.
In the diagonalized $W^\pm$ sector, we get
\begin{eqnarray}
-i\Pi_{W^\pm f}^L&=&-\frac{ig^2\Lambda_W^2}{(4\pi)^2}\sum_{q^L}(K_{m_1m_2}-L_{m_1m_2}),\\
-i\Pi_{W^\pm f}^T&=&-\frac{ig^2\Lambda_W^2}{(4\pi)^2}\sum_{q^L}(K_{m_1m_2}-L_{m_1m_2}+2P_{m_1m_2}).
\label{eq:WWPi}
\end{eqnarray}

We note that at $p^2=0$,
\begin{equation}
-i\Pi_{W^\pm f}^L\bigg|_{p^2=0}=-i\Pi_{W^\pm f}^T\bigg|_{p^2=0}=-\frac{ig^2\Lambda_W^2}{(4\pi)^2}\sum_{q^L}(K_{m_1m_2}-L_{m_1m_2})
\bigg|_{p^2=0}\ne 0.
\end{equation}
This introduces three Nambu-Goldstone degrees of freedom into the $W-B$ sector and the vector bosons W$^\pm$ and Z$^0$ acquire longitudinal parts and corresponding masses. We observe that $\Pi_A^T(0)=0$, guaranteeing a massless photon \cite{Moffat2008b}.

\section{Applications}
\label{sec:appl}

Comparing (\ref{eq:vecprop}) to the standard model vector boson propagator allows us to calculate the masses of the $W^\pm$ and $Z^0$ bosons or conversely, use their experimentally known masses to calculate $\Lambda_W$. On the other hand, the fermion self-energy graphs allow us to generate fermion masses. The coupling constants of the theory are shown to be running; this plays an important role in guaranteeing unitary behavior.

\subsection{Calculation of the $\rho$ parameter and $\Lambda_W$}

When we consider the scattering of longitudinally polarized vector bosons, the vector boson propagator (\ref{eq:vecprop}) reads
\begin{equation}
iD^{\mu\nu}(p^2)=\frac{-i\eta^{\mu\nu}}{p^2-\Pi_f^T(p^2)},
\end{equation}
where we explicitly indicated the dependence of the self-energy and the propagator on momentum. This differs from the vector boson propagator of the standard model in that the squared mass $m_V^2$ of the vector boson is replaced by the self-energy term $\Pi_f^T$. For an on-shell vector boson, demanding agreement with the standard model requires that the following consistency equation be satisfied:
\begin{eqnarray}
&&~~~~m_V^2=\Pi_f^T(m_V^2),\\
&&-i\Pi_{Zf}^T=-\frac{1}{2}\frac{i(g^2+g'^2)\Lambda_W^2}{(4\pi)^2}\nonumber\\
&&\times\sum_\psi[(K_{mm}-L_{mm})+P_{mm}(2c_w^4+s_w^432(Q-T^3)^2-16s_w^2c_w^2T^3(Q-T^3))].~~~~~~
\label{eq:ZZPi}
\end{eqnarray}

This equation contains terms that include the electroweak coupling constant, the Weinberg angle, fermion masses, and the $\Lambda_W$ parameter. As all these except $\Lambda_W$ are known from experiment, the equation
\begin{equation}
m_Z^2=\Pi_{Zf}^T(m_Z^2),
\end{equation}
the right-hand side of which contains $\Lambda_W$, can be used to determine $\Lambda_W$. Using $g\simeq 0.649$, $\sin^2\theta_w\simeq 0.2312$, $m_t\simeq 171.2$~GeV, $m_Z\simeq 91.1876$~GeV, we get
\begin{equation}
\Lambda_W\simeq 541.9~\mathrm{GeV}.
\end{equation}

Knowing $\Lambda_W$ allows us to solve the consistency equation for the W boson mass. Treating $m_W$ as unknown, we solve using
\begin{equation}
-i\Pi_{W^\pm f}^T=-\frac{ig^2\Lambda_W^2}{(4\pi)^2}\sum_{q^L}(K_{m_1m_2}-L_{m_1m_2}+2P_{m_1m_2}),
\end{equation}
and obtain \cite{Moffat2008b}:
\begin{equation}
m_W\simeq 80.05~\mathrm{GeV}.
\end{equation}
This result, which does not incorporate radiative corrections, is actually slightly closer to the experimental value $m_W=80.398\pm 0.025$~GeV than the comparable tree-level standard model prediction $m_W=79.95$~GeV, obtained using $\rho=1$ where
\begin{equation}
\rho=\frac{m_W^2}{m_Z^2c_w^2}.
\end{equation}
We get, from our model,
\begin{equation}
\rho\simeq 1.0023,
\end{equation}
which agrees well with estimates from the experimental ratio of neutral to charged currents.

It is anticipated that our result for $m_W$ (correct to 0.5\%) will get closer to the experimental value when radiative corrections are included, for our regularization scheme will introduce some suppression of higher-order corrections at the energy scale of $m_W$.

\subsection{Fermion masses}

We will generate fermion masses from the finite one-loop fermion self-energy graph:
\vskip 6pt
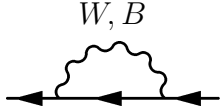
\begin{center}
\begin{fmfgraph*}(80,60)
\fmfleftn{i}{1}\fmfrightn{o}{1}
\fmf{fermion}{v1,i1}
\fmf{fermion}{o1,v2}
\fmf{fermion,tension=0.5}{v2,v1}
\fmf{boson,right,tension=0,label=$W,,B$}{v2,v1}
\end{fmfgraph*}
\end{center}
\vskip -30pt
This method of deriving fermion masses is more economical in assumptions, as we obtain the masses from our original massless electroweak Lagrangian by calculating fermion self-energy graphs \cite{Moffat2007f,Moffat2008b}.

A fermion particle obeys the equation
\begin{equation}
\slashed p-m_{0f}-\Sigma(p)=0,~~~~~\slashed p-m_f=0.
\end{equation}
Here, $m_{0f}$ is the bare fermion mass, $m_f$ is the observed fermion mass and $\Sigma(p)$ is the finite proper self-energy part. We have
\begin{equation}
m_f-m_{0f}=\Sigma(p,m_f,g,\Lambda_f)\big|_{\slashed p-m_f=0},
\end{equation}
where $\Lambda_f$ denotes the energy scales for lepton and quark masses.

A solution can be found by successive approximations starting from the bare mass $m_{0f}$.

The one-loop correction to the self-energy of a fermion with mass $m_f$ in the regularized theory for a massive vector field is obtained from $\Sigma(p)$.

We now identify the fermion mass as $m_f=\Sigma(0)$:
\begin{equation}
m_f=\frac{g^2}{4\pi^2}\exp\left(\frac{-m_V^2}{\Lambda_f^2}\right)m_f\left[E_1\left(\frac{2m_f^2}{\Lambda_f^2}\right)-\frac{m_V^2}{\Lambda_f^2}\int\limits_2^\infty d\tau\exp\left(\tau\frac{m_V^2-m_f^2}{\Lambda_f^2}\right)E_1\left(\tau\frac{m_V^2}{\Lambda_f^2}\right)\right].
\label{eq:mf}
\end{equation}

In addition to admitting a trivial solution at $m_f=0$, this equation also has non-trivial solutions that can be computed numerically. In a theory with a single massless vector boson, we get
\begin{equation}
m_f=\frac{g^2}{4\pi^2}m_fE_1\left(\frac{2m_f^2}{\Lambda_f^2}\right),~~~~~\frac{m_f}{\Lambda_f}=\sqrt{\frac{1}{2}E^{-1}_1\left(\frac{4\pi^2}{g^2}\right)}.
\end{equation}
For a top quark mass $m_t=171.2$~GeV, the corresponding energy scale is about $\Lambda_t\simeq 6$~TeV.

In these calculations, $\Lambda_f$ plays a role that is similar to that of the diagonalized fermion mass matrix in the standard model. The number of undetermined parameters, therefore, is the same as in the standard model: for each fermion a corresponding $\Lambda_f$ determines its mass.

\subsection{The running of coupling constants and unitarity}

The Higgs field resolves the issue of unitarity by {\bf precisely} canceling out badly behaved terms in the tree-level amplitude of processes involving longitudinally polarized vector bosons, for instance, W$^+_L$W$^-_L\rightarrow$W$^+_L$W$^-_L$ or e$^+$e$^-\rightarrow$W$^+_L$W$^-_L$. The challenge to any theory that aims to compete with the SM without introducing a Higgs particle is to generate the correct fermion and boson masses on the one hand, and ensure unitary behavior for these types of scattering processes on the other \cite{Moffat2008c}.

Given the way $\Pi^T$ appears in the vector boson propagator, it is reasonable to make the identification:
\begin{equation}
\Pi_{Wf}^T(q^2)=m_W^2(q^2),~~~~~\Pi_{Zf}^T(q^2)=m_Z^2(q^2).\label{eq:PIWZ}
\end{equation}

When we rewrite the theory's Lagrangian in terms of massive vector bosons, the Lagrangian picks up a finite mass contribution from the total sum of polarization graphs:
\begin{eqnarray}
\label{massmatrix}
L_m&=&\frac{1}{8}v^2g^2[(W^1_\mu)^2+(W^2_\mu)^2]+\frac{1}{8}v^2[g^2(W^3_\mu)^2-2gg'W^3_\mu B^\mu+g^{'2}B^2_\mu]\nonumber\\
&=&\frac{1}{4}g^2v^2W^+_\mu W^{-\mu}+\frac{1}{8}v^2(W_{3\mu},B_\mu)\left(\matrix{g^2 &
-gg'\cr-gg' &
g^{'2}}\right)\left(\matrix{W^{3\mu}\cr B^\mu}\right),\\
&&m_W=\frac{1}{2}vg,~~~~~m_Z=\frac{1}{2}v\sqrt{g^2+g'^2},~~~~~m_A=0,
\end{eqnarray}
where $v\propto\Pi$ is the electroweak symmetry breaking scale (which, in the SM, is the vacuum expectation value of the Higgs scalar).

Consistency requires the running of the constants $g$ and $g'$. Starting with the W mass, we obtain
\begin{equation}
\frac{g^2(q^2)}{g^2(m_Z^2)}=\frac{\Pi_{Wf}^T(q^2)}{\Pi_{Wf}^T(m_Z^2)},~~~~~v^2=\frac{4\Pi_{Wf}^T(m_Z^2)}{g^2(m_Z^2)}\simeq (245~\mathrm{GeV})^2.
\end{equation}
Using the Z mass we get
\begin{equation}
\frac{g^2(q^2)+g'^2(q^2)}{g^2(m_Z^2)+g'^2(m_Z^2)}=\frac{\Pi_{Zf}^T(q^2)}{\Pi_{Zf}^T(m_Z^2)},
\end{equation}
which establishes the running of $g'(q^2)$. These relationships also allow us to calculate the running of the Weinberg angle $\theta_w$, which is defined through the ratio of the coupling constants $g$ and $g'$ as
\begin{equation}
\cos\theta_w=\frac{\sqrt{g^2+g'^2}}{g}.
\end{equation}

In the high energy limit, electroweak yields the matrix element for $\mathrm{e}^+\mathrm{e}^-\rightarrow\mathrm{W}_L^+\mathrm{W}_L^-$:
\begin{equation}
i{\cal M}=-ig^2\left[\frac{m_e}{2m_W^2}\sqrt{s}+{\cal O}(1)\right],
\end{equation}
which, in the Standard Model, is canceled out by the Higgs exchange. Similarly, for $\mathrm{W}_L^+\mathrm{W}_L^-\rightarrow\mathrm{W}_L^+\mathrm{W}_L^-$, electroweak theory yields
\begin{equation}
i{\cal M}=ig^2\left[\frac{\cos\theta+1}{8m_W^2}s+{\cal O}(1)\right],
\end{equation}
which is again canceled out by the Higgs exchange. In the case of the Higgless FEW theory, no such additive cancelations takes place. However, the running of the electroweak coupling constant is such that at high $s$, $g(s)s\sim$~const., which is sufficient to ensure that unitarity is not violated \cite{Moffat2008c}.

\section{Conclusions}

An electroweak model without a Higgs particle that breaks $\mathrm{SU}(2)_L\times\mathrm{U}(1)_Y$ symmetry has been developed, based on a finite quantum field theory. We begin with a massless and gauge invariant theory that is UV complete, Poincar\'e invariant and unitary to all orders of perturbation theory. A fundamental energy scale $\Lambda_W$ enters into the calculations of the finite Feynman loop diagrams. A path integral is formulated that generates all the Feynman diagrams in the theory. The self-energy boson loop graphs with internal fermions comprising the observed 12 quarks and leptons have an associated measure in the path integral that is broken to generate 3 Nambu-Goldstone scalar modes that give the W$^\pm$ and the Z$^0$ bosons masses, while retaining a zero mass photon.

{\bf There is no classical Higgs scalar field particle and no new particles are included in the particle spectrum. All particle masses are generated by QFT self-energy diagrams.}

The W$^+_L$W$^-_L\rightarrow$W$^+_L$W$^-_L$ or e$^+$e$^-\rightarrow$W$^+_L$W$^-_L$ amplitudes do not
violate unitarity at the tree graph level due to the running with energy of the electroweak coupling constants $g$, $g'$ and $e$. This is essential for the physical consistency of the model as is the
case in the standard Higgs electroweak model.

A self-consistent calculation of the energy scale yields $\Lambda_W=542$~GeV and a prediction of the W mass from the W boson self-energy diagrams in the symmetry broken phase gives $m_W=80.05$~GeV, which is accurate to 0.5\%.

The EW cosmological constant problem and the Higgs mass hierarchy problem are both solved without fine-tuning. The origin of mass in the universe is due to self-consistent solutions of QFT self-energies---not to a classical scalar Higgs field and Yukawa interactions.

\end{fmffile}

\bigskip
\bigskip
\bigskip

The research was partially supported by National Research Council of Canada. Research at the Perimeter Institute for Theoretical Physics is supported by the Government of Canada through NSERC and by the Province of Ontario through the Ministry of Research and Innovation (MRI).

\bibliographystyle{unsrt}
\bibliography{refs}

%
%
%
%
%
%
%
%

\end{document}